\documentclass[conference]{IEEEtran}

\usepackage{times,epsfig}
\usepackage{epsf}
\usepackage{graphicx}
\include{psfig}
\include{epsf}
\include{epsfrot}
\usepackage{amsmath}
\usepackage{amssymb}
\usepackage{latexsym}
\usepackage{stfloats}
\usepackage{citesort}
\usepackage{fancyhdr}
\usepackage{stmaryrd}

\usepackage{tikz}
\usepackage{lipsum}

\renewcommand{\baselinestretch}{.97}

\def\cE{{\mathcal{E}}}

\def\b0{{\mathbf{0}}}

\newcommand{\bet}{\begin{table}}
\newcommand{\eet}{\end{table}}
\newcommand{\btt}{\begin{tabular}}
\newcommand{\ett}{\end{tabular}}
\newcommand{\bec}{\begin{center}}
\newcommand{\eec}{\end{center}}
\newcommand{\bef}{\begin{figure}}
\newcommand{\eef}{\end{figure}}
\newcommand{\beq}{\begin{eqnarray}}
\newcommand{\eeq}{\end{eqnarray}}
\newcommand{\beqs}{\vspace*{-0.05in}\begin{eqnarray}}
\newcommand{\eeqs}{\end{eqnarray}\vspace*{-0.05in}}
\newcommand{\bit}{\begin{itemize}}
\newcommand{\eit}{\end{itemize}}
\newcommand{\bed}{\begin{description}}
\newcommand{\eed}{\end{description}}
\newcommand{\ben}{\begin{enumerate}}
\newcommand{\een}{\end{enumerate}}
\newcommand{\bis}{\vspace*{-0.05in}\begin{itemize}\small}
\newcommand{\eis}{\end{itemize}\normalsize\vspace*{-0.05in}}

\newcommand{\fig}[1]{Fig.\ \ref{#1}}

\newcommand{\tightldots}{...}

\def\-{\! - \!}
\def\+{\! + \!}
\def\={\! = \!}
\def\>{\! > \!}

\makeatletter
\def\@oddfoot{\hbox{}\hfil \scriptsize  Copyright 2015 SS\&C. Submitted to the IEEE Asilomar Conference on Signals, Systems, and Computers, Nov. 8-11$^{\rm th}$ 2015, Pacific Grove, CA, USA \hfill}
\newcommand{\ps@edbfooter}{
   \renewcommand{\@oddfoot}{\hbox{}\hfil \scriptsize  Copyright 2015 SS\&C. Submitted to the IEEE Asilomar Conference on Signals, Systems, and Computers, Nov. 8-11$^{\rm th}$ 2015, Pacific Grove, CA, USA \hfill}    
}
\makeatother

\begin{document}

\title{Achieving Large Multiplexing Gain\\
in Distributed Antenna Systems \\
via Cooperation with pCell Technology}

\author{
Antonio Forenza$^{*}$, Stephen Perlman$^{*}$, Fadi Saibi$^{*}$, Mario Di Dio$^{*}$, Roger van der Laan$^{*}$, Giuseppe Caire${}^{\dagger}$\\
[-7pt] ${}$\\
[-2pt] \small $^{*}$Artemis Networks, LLC \\
[-2pt] \small 355 Bryant Street, Suite 110\\
[-2pt] \small San Francisco, CA, 94107, USA\\
[-7pt] ${}$\\
[-2pt] \small ${}^{\dagger}$Technische Universit\"{a}t Berlin\\
[-2pt] \small Communications and Information Theory Chair\\
[-2pt] \small Einsteinufer 25, 10587 Berlin, Germany}

\maketitle


\begin{abstract}
In this paper we present pCell\textsuperscript{TM} technology, the first commercial-grade wireless system that employs cooperation between distributed transceiver stations to create concurrent data links to multiple users in the same spectrum. First we analyze the per-user signal-to-interference-plus-noise ratio (SINR) employing a geometrical spatial channel model to define volumes in space of coherent signal around user antennas (or personal cells, i.e., pCells). Then we describe the system architecture consisting of a general-purpose-processor (GPP) based software-defined radio (SDR) wireless platform implementing a real-time LTE protocol stack to communicate with off-the-shelf LTE devices. Finally we present experimental results demonstrating up to 16 concurrent spatial channels for an aggregate average spectral efficiency of 59.3 bps/Hz in the downlink and 27.5 bps/Hz in the uplink, providing data rates of 200 Mbps downlink and 25 Mbps uplink in 5 MHz of TDD spectrum.\footnote{The work of the Artemis team was supported by Rearden, LLC.\\
\textcopyright 2015 IEEE. Personal use of this material is permitted. Permission from IEEE must be obtained for all other uses, in any current or future media, including reprinting/republishing this material for advertising or promotional purposes, creating new collective works, for resale or redistribution to servers or lists, or reuse of any copyrighted component of this work in other works.}
\end{abstract}

\section{Introduction}
\label{sec_Introduction}

The increasing popularity of smartphones and tablets, and the growing demand for data-hungry applications like HD video streaming has resulted in skyrocketing mobile data traffic. A recent report by the CTIA to the FCC showed mobile data traffic will continue to grow throughout the next four years at an annual rate of about $40\%$ \cite{CTIA-2015}, and states that cellular densification (through small-cells in 4G LTE networks) will be unable to keep pace with this growing demand for more data within current spectrum. New spectrum allocation may be a short-term fix, but mobile spectrum is finite while data demand will grow indefinitely. One solution is to radically improve the spectral efficiency (SE) of wireless networks. In this paper we present pCell, a new wireless technology capable of achieving SE over an order of magnitude higher than any current technology while remaining compatible with existing 4G LTE devices \cite{Artemis-2015}. pCell achieves these gains by forgoing cellularization and exploiting interference in wireless networks through large-scale cooperation between distributed transceivers, and by enabling high spatial multiplexing gain via multiuser transmissions.

A multiuser wireless system with multiple transceiver stations in its simplest form consists of $N$ transmit antennas and $U$ single-antenna receivers (users), which in prior literature is referred to as multiuser multiple-input multiple-output (MU-MIMO). The information theoretic model underlying MU-MIMO is the so-called Gaussian vector broadcast channel, and has been the subject of intense investigation 
started with the work of Caire and Shamai \cite{Caire-Shamai-TIT03}, that found the sum capacity for the case $U = 2$ and proposed linear beamforming with
interference pre-cancellation, known as dirty-paper coding \cite{Costa-TIT83}, as a general achievability strategy. 
Successively,  the sum capacity for general $U \geq 2$ was found almost simultaneously in \cite{Viswanath-Tse-TIT03,Vishwanath-Jindal-Goldsmith-TIT03,Yu-Cioffi-TIT04}. The full characterization of
the capacity region (with no common message) was eventually given in \cite{Weingarten-Steinberg-Shamai-TIT06}, where the optimality of beamforming and dirty-paper coding was shown for 
a general convex input covariance constraint. These results assume that the channel state information (CSI) is fixed and fully known to the transmitter and to the receivers. The extension of the above results to the 
case where the CSI is a random fading matrix, also known to all instantaneously, are almost immediate, especially for the case of ergodic rates, i.e., when the CSI evolves over time according to a matrix stationary and ergodic process. 

Moving from theoretical models to real-world systems, several practical limitations arise. Given the low spatial diversity yielded by centralized antenna structures, performance of MIMO (or MU-MIMO) systems mostly relies upon the limited multi-paths available in propagation channels \cite{3GPP-SCM-2003,TGnChannelMod-2004}, and in practice at most 4x spatial multiplexing gain is achieved \cite{Ericsson-2013,Quantenna2-2009}. One solution is to utilize far more antennas than the number of users to increase spatial diversity, as in massive MIMO systems \cite{Marzetta-TWC10,Larsson-2014}, and create independent spatial channels to multiple concurrent users via beamforming techniques. Massive MIMO, however, relies on highly complex base station designs with many tightly-packed RF chains and a centralized antenna architecture which still limits the degrees of freedom in wireless channels. 

The benefit of a de-centralized cellular architecture was studied in \cite{huh2012achieving} showing that spectral efficiency comparable to massive MIMO systems can be achieved with one order of magnitude fewer antennas via network MIMO, by enabling cooperation between base stations in adjacent cells to mitigate inter-cell interference for enhancing cell-edge performance of cellular systems \cite{NetMIMO-Foschini-2006,NetMIMO-Karakayali-2006,NetMIMO-Karakayali2-2006,NetMIMO-Liang-2007,NetMIMO-Venkatesan-2007,NetMIMO-Venkatesan-2009,NetMIMO-Lozano-2013}. From the theoretical viewpoint, network MIMO is equivalent to the Gaussian vector broadcast channel reviewed before \cite{Somekh-Shamai-TIT07,huh2012network}, 
unless one takes explicitly into account the constraints imposed by the underlying wired network that connects the remote radio heads (RRHs)
to the central processor. This network is referred to as backhaul, when the RRHs are seen as individual base stations that somehow cooperate (in the so-called CoMP schemes) or, more modernly, as fronthaul, when the RRHs are simple and relatively dumb devices that form a whole distributed base station together with the central processor (as in a C-RAN architecture). 
Depending on the constraints on the backhaul/fronthaul (e.g., topology \cite{annapureddy2012degrees}, link rates \cite{hong2013compute}) and type of cooperation (e.g., full joint processing \cite{Somekh-Shamai-TIT07}, 
coordinated beamforming \cite{Huh-Caire-TSP10},  interference avoidance \cite{Caire-Docomo-Allerton08}), a very large number of  information theoretic problems and corresponding schemes 
have been investigated in the literature. 

There are fundamental capacity limits in cooperative networks operating within the cellular framework \cite{NetMIMO-Lozano-2013}, where the spectral efficiency reaches an upper limit due to out-of-cluster interference overwhelming the in-cluster signals, as transmit power increases. In this paper we propose a different network architecture with transceivers distributed serendipitously without any concept of a cell, exploiting high densification with fixed transmit power to increase spatial multiplexing gain. The transceivers are connected through a fronthaul and cooperate on a large scale to create concurrent spatial channels to multiple users via precoding.

We begin by showing the benefits of a distributed architecture over conventional cellular networks with multiple centralized antennas, through a geometrical propagation model in Section \ref{sec_Analysis}. We use this model to analyze the signal-to-interference-plus-noise ratio (SINR) as a function of system parameters and demonstrate that by distributing the transceivers randomly in space it is possible to achieve volumes of coherent signal  with high SINR around every user antenna, which is impractical in centralized antenna systems. Section \ref{sec_SDR} describes the GPP-based SDR wireless platform implementing in real-time the pCell processing and the entire LTE protocol stack. Finally, Section \ref{expResults} demonstrates experimentally how the volumes of coherent signal enable high multiplexing gain in a practical propagation environment. 


\section{System Model and Analysis}
\label{sec_Analysis}

\subsection{System and Channel Model}
\label{subsec_ChanModel}

The downlink of a multiuser system is modeled for a single channel use on the time-frequency plane as\footnote{We use ${}^*$ to denote conjugation, ${}^T$ to denote transposition, ${}^\dagger$ to denote conjugation and transposition, $|\cdot|$ to denote the absolute value, $||\cdot||$ to denote the 2-norm, ${\bf I}_{U}$ to denote the identity matrix of size $U \times U$, ${\mathbb C}^{U \times N}$ to denote complex matrix of size $U \times N$, $\left<\cdot,\cdot\right>$ to denote the complex vector space inner-product, $\odot$ to denote element-by-element vector multiplication, $\vec{\cdot}$ to indicate a vector in the 3-dimensional physical space, $\hat{\cdot}$ to denote such vector of unit-norm, and $i$ is the imaginary unit.} 
\begin{equation}
    \label{eqRxSignal}
    {\bf y} =  {\bf H} {\bf x} + {\bf n}
\end{equation}
where ${\bf y} \in {\mathbb C}^{U \times 1}$ is the receive
signal vector, ${\bf x} \in {\mathbb C}^{N \times 1}$ is the transmit signal vector subject to the power constraint
$\cE\{||{\bf x}||^2\} = U$, ${\bf n} \in {\mathbb C}^{U \times 1}$ is the zero-mean additive white Gaussian noise vector with
covariance matrix $\cE\{{\bf n}{\bf n}^\dagger\} = N_{\rm o} {\bf
I}_{U}$, and ${\bf H} \in {\mathbb C}^{U \times N}$ is the channel matrix with $N \geq U$, describing the propagation from the $N$ transmit antennas to the $U$ receive antennas. 
The channel vector ${\bf h}_u = \left[h_{u1}, \ldots, h_{uN}\right]^{\dagger} \in {\mathbb C}^{N \times 1}$ is associated with each user $u$ such that the channel 
matrix is given by ${\bf H} = \left[{\bf h}_1,\dots,{\bf h}_U\right]^\dagger \in {\mathbb C}^{U \times N}$.

\begin{figure} [!h]  
	\centerline{\resizebox{0.81\columnwidth}{!}{\includegraphics{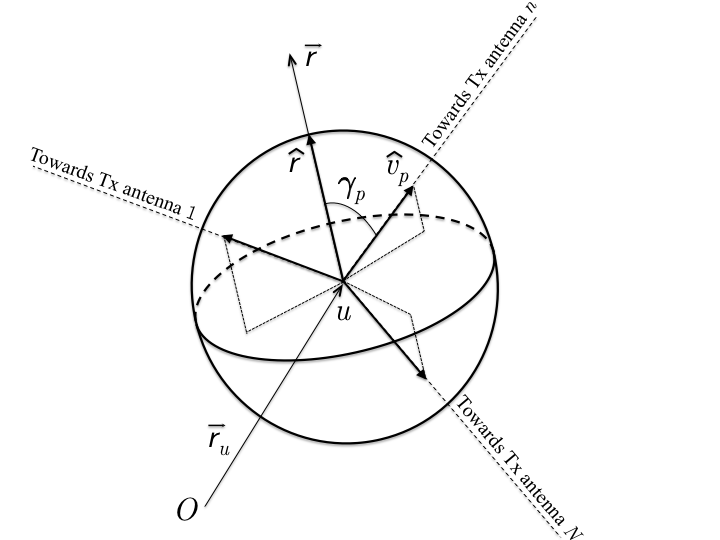}}}
	\caption{Model parameters for LOS channels: angle $\gamma_{p} \in [0,\pi]$ between path direction $\hat v_{p}$ and displacement direction $\hat r$ with $N$ transmit antennas.} 
	\label{fig:displacementCosGammas}
\end{figure}

We use a channel model that accounts for the spatial dependency of the electromagnetic field between the transmit and receive antennas. Through this model we define the notion of volumes in space of coherent signal and investigate how system configuration parameters affect its geometry.
For the sake of simplicity, we assume point antennas that are unpolarized and isotropic radiators in far-field.
We use the model of spherical waves in scattering environments as in \cite{Svantesson-2000}. For ease of exposition, we also make the assumption that the distance between users is much smaller than their distance to transmitters and scatterers as in \cite{poonTse2005} so that the complex channel coefficient between transmit antenna $n = 1,\tightldots,N $ and receive antenna $u = 1,\tightldots,U$ is modeled as a superposition of plane waves
\begin{equation}
\label{channelCoefficientModel}
h_{un} = \sum_{p \in \mathcal{S}_{un}} a_p e^{-i k \hat v_p \cdot \vec r_u}
\end{equation}
where $\mathcal{S}_{un}$ is the set of all paths, including line-of-sight (LOS) and non-line-of-sight (NLOS) components\footnote{Such that $a_{p}$ includes the term $\sqrt{K/(K+1)}$ (where $K$ is the Rician K-factor) for the LOS component and  $\sqrt{1/(K+1)}$ for the NLOS components.}, from propagation and cluster scattering from transmit antenna $n$ to receive antenna $u$, $k = 2\pi / \lambda$ is the wavenumber, $\lambda$ is the wavelength, $\vec r_u$ is the location vector of user $u$ relative to an origin $O$ as shown in \fig{fig:displacementCosGammas}, $\hat v_p$ is the unit vector in the direction of the incident path $p$ pointing out from the location of user $u$, and $a_p$ is a complex coefficient modeling pathloss, shadowing and phase terms independent of $\vec r_u$ for path $p$.

\subsection{SINR Performance via Multipole Expansion}
\label{subsec_Sinr}

The system performs transmit precoding to create multiple independent downlink data streams to the users.
In general the transmit precoding is adaptively adjusted based on CSI of the users.
When the transmit precoding is fixed and the user at the location $\vec r_u$ is displaced by $\vec r$, the received SINR changes from $\textrm{SINR}_u$ at $\vec r = \vec o$ to $\textrm{SINR}_u(\vec r)$ as the user's channel varies from ${\bf h}_u$ to ${\bf h}_u(\vec r)$.
 We define the volume of coherent signal for user $u$ as the space where $\textrm{SINR}(\vec r)$ at the displaced user's location exceeds a threshold $\textrm{SINR}_{\rm o}$
\begin{equation}
  \label{pCellDef}
  \mathcal{V}_u(\textrm{SINR}_{\rm o}) = \lbrace\vec r_u + \vec r; \, \textrm{SINR}_u(\vec r) \geq \textrm{SINR}_{\rm o}\rbrace
\end{equation}
where $\textrm{SINR}_{\rm o}$ is chosen to meet a predefined error rate or capacity performance.
To derive an expression of the SINR$(\vec r)$ that allows insightful review through a tractable analytical form, we assume ${\bf x} = {\bf W} {\bf s}$ in (\ref{eqRxSignal}), where ${\bf s} \in {\mathbb C}^{U \times 1}$ is the transmit signal vector with power constraint $\cE\{||{\bf s}||^2\} = U$ and ${\bf W} = \left[{\bf w}_1, \dots, {\bf w}_U\right] \in {\mathbb C}^{N \times U}$ is a linear zero-forcing precoder as described in \cite{spencerSwindlehurst2004,chenHeath2007}, although other precoding techniques may be applied in practical deployments. 

Assuming the channel matrix ${\bf H}$ is full-rank and all users are allocated equal power, for a given user $u$ displaced by $\vec r$ relative to its original location $\vec r_u$ the SINR is given by
\begin{equation}
  \label{sinrBase}
  \textrm{SINR}_u(\vec r) = \frac{\left| \left< {\bf h}_u(\vec r) , {\bf w}_u \right> \right|^2}{N_{\rm o} + \sum\limits_{v \neq u}\left| \left< {\bf h}_u(\vec r) , {\bf w}_v \right> \right|^2}
\end{equation}
where ${\bf h}_u(\vec r) = \left[h_{u1}(\vec r),\dots, h_{uN}(\vec r)\right]^{\dagger} \in {\mathbb C}^{N \times 1}$ and the entries of the vector are defined from (\ref{channelCoefficientModel}) as $h_{un}(\vec r) = \sum_{p \in \mathcal{S}_{un}} a_p e^{-i k \hat v_p \cdot (\vec r_u + \vec r)}$. Applying the multipole expansion for plane waves \cite{poonTse2005} to the phasor term $ e^{-i k \hat v_p \cdot \vec r}$ yields
\begin{equation}
  \label{channelMultipole}
  {\bf h}_u(\vec r) = \sum_{\ell=0}^{+\infty}i^\ell \, (2\ell+1) \, j_\ell(kr) \, {\bf b}_u^\ell(\hat r) 
\end{equation}
where the displacement vector $\vec r$ is decomposed into its norm $r$ and unit direction vector $\hat r = \vec r / r$, $j_\ell(\cdot)$ is the spherical Bessel function of the first kind and order $\ell$ and ${\bf b}_u^\ell(\hat r) = \left[b_{u1}^\ell(\hat r),\dots,b_{uN}^\ell(\hat r)\right]^{\dagger} \in {\mathbb C}^{N \times 1}$ is such that
\begin{equation}
  \label{eq_Legendre}
  b_{un}^\ell(\hat r) = \sum_{p \in \mathcal{S}_{un}}a_p e^{-i k \hat v_p \cdot \vec r_u} P_\ell(\cos\gamma_p)
\end{equation}
where $P_\ell(\cdot)$ is the Legendre polynomial of degree $\ell$ and $\cos\gamma_p = \hat r \cdot \hat v_p$.
Substituting (\ref{channelMultipole}) into (\ref{sinrBase}) we obtain
\begin{equation}
  \label{sinrMultipole}
  \textrm{SINR}_u(\vec r) = \frac{\left| \sum\limits_{\ell=0}^{+\infty}(-i)^\ell(2\ell+1) j_\ell(kr) \left<{\bf b}_u^\ell(\hat r) , {\bf w}_u \right> \right|^2}{N_{\rm o} + \sum\limits_{v \neq u} \left| \sum\limits_{\ell=1}^{+\infty}(-i)^\ell(2\ell+1) j_\ell(kr) \left< {\bf b}_u^\ell(\hat r) , {\bf w}_v \right> \right|^2}
\end{equation}
where the term of order $0$ in the series at the denominator vanishes since ${\bf b}_u^0(\hat r) = {\bf h}_u$ and we assumed a zero-forcing precoder\footnote{Note that at the user antenna original location (i.e., $\vec r = \vec o$) all terms with $\ell \geq 1$ are zero, $j_0(0) = 1$ and the interference term at the denominator vanishes such that $\textrm{SINR}_u(\vec o) = \textrm{SNR}_u = \left| \left< {\bf h}_u , {\bf w}_u \right> \right|^2\!/N_{\rm o}$.}. 

The expansion in (\ref{channelMultipole}) decouples the effect of displacement distance $r$ and displacement direction $\hat r$, thereby showing how the SINR in (\ref{sinrMultipole}) varies as the user is displaced from its original location.
For example, in the special case of all incident path directions located on a cone with axis being the displacement direction $\hat r$ (that is to say the terms $\cos\gamma_p$ in (\ref{eq_Legendre}) are all equal to a fixed $\cos\gamma$), then ${\bf b}_u^\ell(\hat r) = P_\ell(\cos\gamma) {\bf h}_u$ for any $\ell$.
Therefore, for all $v \neq u$, the terms at the denominator in (\ref{sinrMultipole}) satisfy the condition $\left< {\bf b}_u^\ell(\hat r) , {\bf w}_v \right> = 0$, which entails there is no interference in that displacement direction $\hat r$.

\subsection{Volumes of Coherent Signal}
\label{subsec_Volumes}

Next, we make a few assumptions to clarify the definition of volume of coherent signal in the special case of LOS channels. For small displacement distances, (\ref{sinrMultipole}) yields the approximation
\begin{equation}
  \label{sinrApprox}
  \textrm{SINR}_u(\vec r) \approx \frac{\left| \left< {\bf h}_u , {\bf w}_u \right> \right|^2}{N_{\rm o} + (k r)^2 \sum\limits_{v \neq u} \left|\left< {\bf b}_u^1(\hat r) , {\bf w}_v \right> \right|^2}
\end{equation}
which stems from $j_\ell(\rho) = \rho^\ell/(2\ell+1)!! \, (1+\mathcal{O}(\rho^2))$ as $\rho \rightarrow 0$ \cite{handbookMathFunctions}.
In the case where the only significant path is the LOS path as illustrated in \fig{fig:displacementCosGammas}, then ${\bf b}_u^1(\hat r) = {\bf h}_u \odot [\cos\gamma_{u1},\ldots,\cos\gamma_{uN}]^T$, where we used the fact that $P_1(\cos\gamma) = \cos\gamma$, and $\gamma_{un}$ is the relative angle between the displacement direction $\hat r$ and the unit vector $\hat v_{un}$ from the location of user $u$ to the location of transmit antenna $n$.
Therefore, the magnitude of the directional component of the interference term explicitly depends on the relative angles $\gamma_{un}$ for a given user location. 

By further assuming that the $U$ channel vectors ${\bf h}_u$ are orthogonal, then the zero-forcing precoder becomes so that ${\bf w}_u = {\bf h}_u/||{\bf h}_u||$ for $u = 1,\tightldots,U$.
If in addition $N=U$ then (\ref{sinrApprox}) simplifies to\footnote{If $N > U$ then the right hand side of (\ref{sinrApproxOrth}) becomes an approximate lower bound for $\textrm{SINR}_u(\vec r)$.}
\begin{equation}
  \label{sinrApproxOrth}
  \textrm{SINR}_u(\vec r) \approx \frac{\textrm{SNR}_u}{1 + (k r)^2 \, \textrm{SNR}_u \!\!\!\!\sum\limits_{1 \leq s < t \leq N} \!\!\!\!\xi_{us} \xi_{ut}\left(\cos\gamma_{us} - \cos\gamma_{ut}\right)^2}
\end{equation}
where $\textrm{SNR}_u = \textrm{SINR}_u(\vec o)= ||{\bf h}_u||^2 / N_{\rm o}$ is the SINR for user $u$ at its original location, $\xi_{un} = |h_{un}|^2/||{\bf h}_u||^2$ is the fraction of the total channel power gain from antenna $n$ and $\cos\gamma_{un} = \hat r \cdot \hat v_{un}$. For a fixed displacement direction $\hat r$, the SINR approximation in (\ref{sinrApproxOrth}) is a Lorentzian function of the displacement distance $r$ with maximum value $\textrm{SNR}_u$ at $r=0$. 

Based on the definition of volume of coherent signal in (\ref{pCellDef}), we consider the surface boundary for $\mathcal{V}_u(\textrm{SINR}_{\rm o})$ where the SINR is equal to a predefined threshold value such that $\textrm{SINR}_u(\vec r) = \textrm{SINR}_{\rm o}$. Then we use the approximation in (\ref{sinrApproxOrth}) to derive a closed-form expression of the radius of volume of coherent signal for user $u$ as a function of the displacement direction $\hat r$ as\footnote{Note that this approximation relies on the orthogonality assumption and is only quantitatively valid if $R_u(\hat r) \ll \lambda/2\pi$ but is useful for drawing qualitative conclusions nonetheless. An expression that does not use that assumption can be derived from (\ref{sinrApprox}).}
\begin{equation}
  \label{pCellRadius}
  R_u(\hat r) \approx \frac{\lambda}{2 \pi \sqrt{\sum\limits_{s < t} \xi_{us} \xi_{ut}\left(\cos\gamma_{us} - \cos\gamma_{ut}\right)^2} }  \sqrt{ \frac{1}{\textrm{SINR}_{\rm o}} - \frac{1}{\textrm{SNR}_u} }.
\end{equation}

From (\ref{pCellRadius}) we derive the following important observations about $R_u(\hat r)$:
\begin{itemize}
\item it is proportional to the wavelength $\lambda$;
\item it decreases as $\textrm{SINR}_{\rm o}$ increases;
\item it depends only on $\textrm{SINR}_{\rm o}$ when $\textrm{SNR}_u \gg \textrm{SINR}_{\rm o}$;
\item it depends, for a given displacement direction $\hat r$, on the layout of the transmit antennas through the angles $\gamma_{un}$;
\item it becomes large when the angles $\gamma_{un}$ become close to each other as in centralized antenna arrays.
\end{itemize}

Equation (\ref{pCellRadius}) provides insights on how system parameters affect the geometry of the volumes of coherent signal.
In particular it indicates that a system with distributed transmit antennas can create a volume of coherent signal with very small radius $R_u(\hat r)$ in all directions. Whereas if the transmit antennas are centralized, the radius becomes larger in all directions and, more specifically, much larger in the directions parallel to the line joining the user location to the center of the distant group of transmit antennas.

\begin{figure}[h]
	\centerline{\resizebox{0.92\columnwidth}{!}{\includegraphics{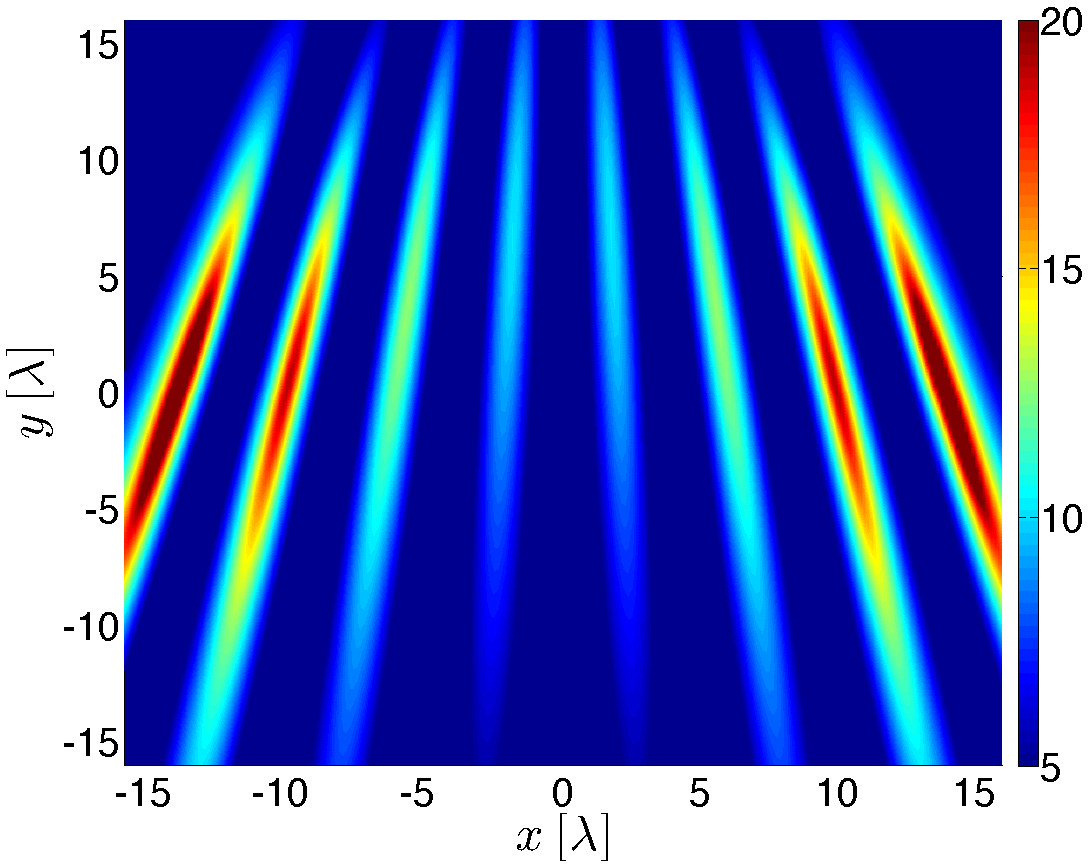}}}
	\caption{Centralized transmit antennas: envelope of $\textrm{SINR}_u(\vec r)$ on the horizontal plane for 8 users regularly spaced $4\lambda$ apart on the axis $\{y=0, \, z=0\}$ and 10 transmit antennas spaced $\lambda/2$ apart on the axis $\{y=50\lambda, \, z=0\}$.} 
	\label{fig:CentralizedULA}
\end{figure}

Indeed for centralized transmit antennas distant from the user locations the terms $(\cos\gamma_{us}-\cos\gamma_{ut})^2$ in (\ref{pCellRadius}) are all small resulting in large dimensions for the volume of coherent signal (if it exists at all\footnote{Note that for centralized antennas the initial assumption of a full-rank channel matrix and even more so the assumption of orthogonal channel vectors is a very hard one to meet in a LOS channel. It requires purposeful placement of the user antennas.
A two-dimensional example is provided in \cite{driessenFoschini1999}.}). 
Furthermore the volume of coherent signal takes on an elongated beam shape.
For example, \fig{fig:CentralizedULA} shows the envelope of the SINR over a two-dimensional cross-section for all users computed through the exact expression in (\ref{sinrBase}) accounting for spherical wave propagation for $U=8$ users uniformly spaced $4\lambda$ apart and placed parallel at the broadside of a $\lambda/2$ uniform linear array (ULA) of $N=10$ transmit antennas located $50\lambda$ away.
The channel model is urban microcell LOS with pathloss and shadowing computed according to the 3GPP model \cite{3GPP-SCM-2003}.
Note that for all users the radius of the volumes of coherent signal is much larger in the direction pointing to the center of the ULA.

\begin{figure} [t]  
	\centerline{\resizebox{0.92\columnwidth}{!}{\includegraphics{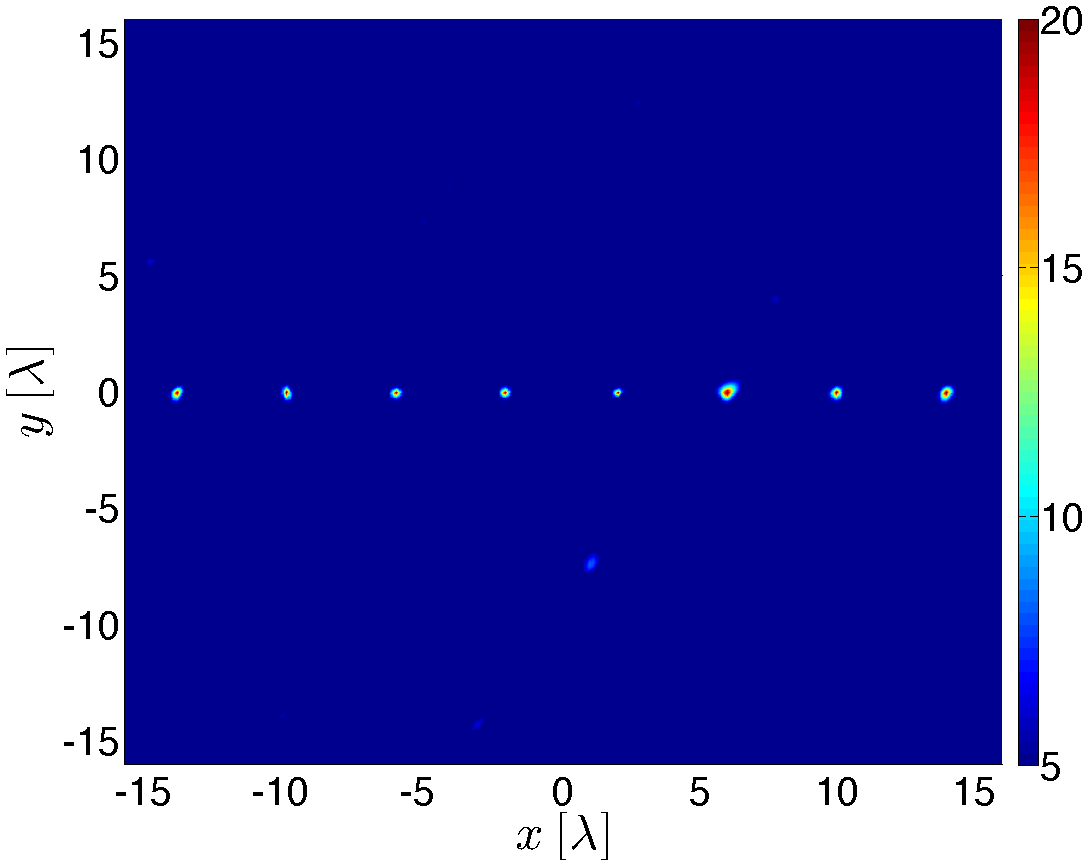}}}
	\caption{Distributed transmit antennas: envelope of $\textrm{SINR}_u(\vec r)$ on the horizontal plane for 8 users  regularly spaced $4\lambda$ apart on the axis $\{y=0, \, z=0\}$ and 10 transmit antennas randomly distributed above the user antennas within the region $\{-50\lambda \leq x \leq +50\lambda, \, -50\lambda \leq y \leq +50\lambda, \, +50\lambda \leq z \leq +200\lambda\}$.} 
	\label{fig:Distributed}
\end{figure}

By contrast, when the transmit antennas are distributed and randomly placed, most of the terms $(\cos\gamma_{us}-\cos\gamma_{ut})^2$ in (\ref{pCellRadius}) are not negligible, thereby the volumes of coherent signal have smaller dimensions in all directions.
For example,  \fig{fig:Distributed} shows the envelope of the SINR for the same conditions as in \fig{fig:CentralizedULA} but for distributed transmit antennas instead.

\begin{figure} [b]  
	\centerline{\resizebox{0.96\columnwidth}{!}{\includegraphics{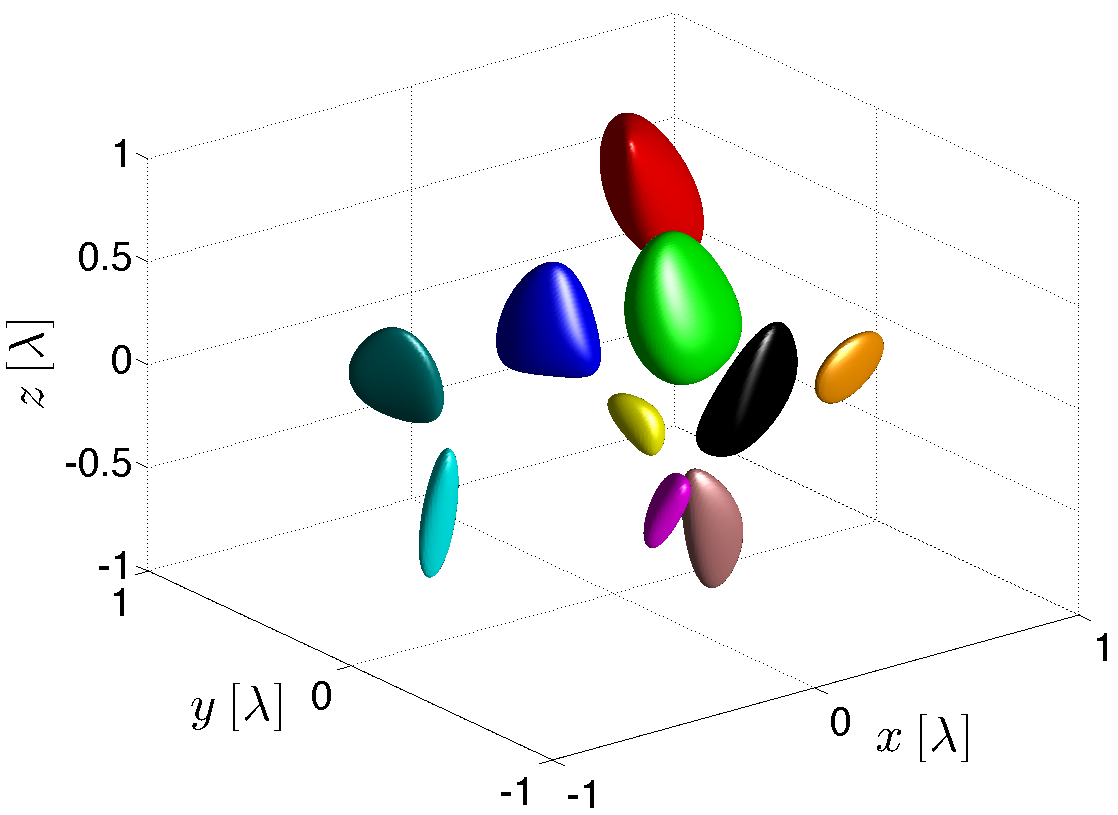}}}
	\caption{Volumes of coherent signal for $U=10$ users randomly placed in a cube of side dimension $2\lambda$ for $N=16$ transmit antennas distributed above the user locations in $\{-300\lambda \leq x \leq +300\lambda, \, -300\lambda \leq y \leq +300\lambda, \, +200\lambda \leq z \leq +300\lambda\}$. Different colors refer to different users.} 
	\label{fig:volCoher}
\end{figure}

\fig{fig:volCoher} shows the volume of coherent signal in (\ref{pCellDef}) around each user with $\textrm{SINR}_{\rm o}=5 {\rm dB}$ (e.g., corresponding to {CQI} 7 or 16-{QAM} with spectral efficiency of 1.48 bps/Hz at a block error rate of 10\% as per the LTE standard \cite{3GPP-CQI-2009,MCS-2010}).
The SINR is computed using (\ref{sinrBase}) for a system with $U=10$ users randomly located in a cube of side dimension $2\lambda$ and $N=16$ transmit antennas distributed above the UE locations with a maximum distance of $300\lambda$ (i.e., realistic dimensions for a deployment at 1.9 GHz carrier frequency).

We obtained experimental evidence of the volumes of coherent signal by measuring variations of the SINR for fixed precoding
while displacing user antennas ever so slightly from their baseline position.
These experiments indicated that the size of the volume of coherent signal is a fraction of the wavelength, such that user devices can be densely packed as in the experimental results in Section \ref{expResults}.
Furthermore, the precoder of the pCell wireless platform is periodically updated such that the volumes of coherent signal follow users' motion. 
Experiments also showed that the volume of coherent signal actually depends on displacement variables $(\vec r, \vec \psi)$ in a 6-dimensional manifold, where in addition to the 3-dimensional variable $\vec r$ used in the present model the variable $\vec \psi$ belongs to the 3-dimensional manifold of Euler angles parameterizing the rotation of each user antenna.
Future work will analyze dependency on rotation angles by extending the present model to account for antenna radiation pattern and polarization.

\psfull

\section{Description of the SDR Wireless Platform}
\label{sec_SDR}

The pCell software-defined radio (SDR) wireless platform is implemented as a cloud radio access network (C-RAN), where baseband processing is performed on GPP servers in a data center. The data center provides I/Q waveforms through fiber connections to RRHs called pWave\textsuperscript{TM} radios, which consist only of analog-to-digital (A/D), digital-to-analog (D/A), and RF up/down converters, power amplifier and antenna. This section describes the hardware and software architectures of the pCell SDR wireless platform as well as aspects of its operation with existing off-the-shelf LTE devices.

\subsection{Hardware Architecture}
\label{subsec_hardwareArchitecture}

\begin{figure} [t]  
        \centerline{\resizebox{1.05\columnwidth}{!}{\includegraphics{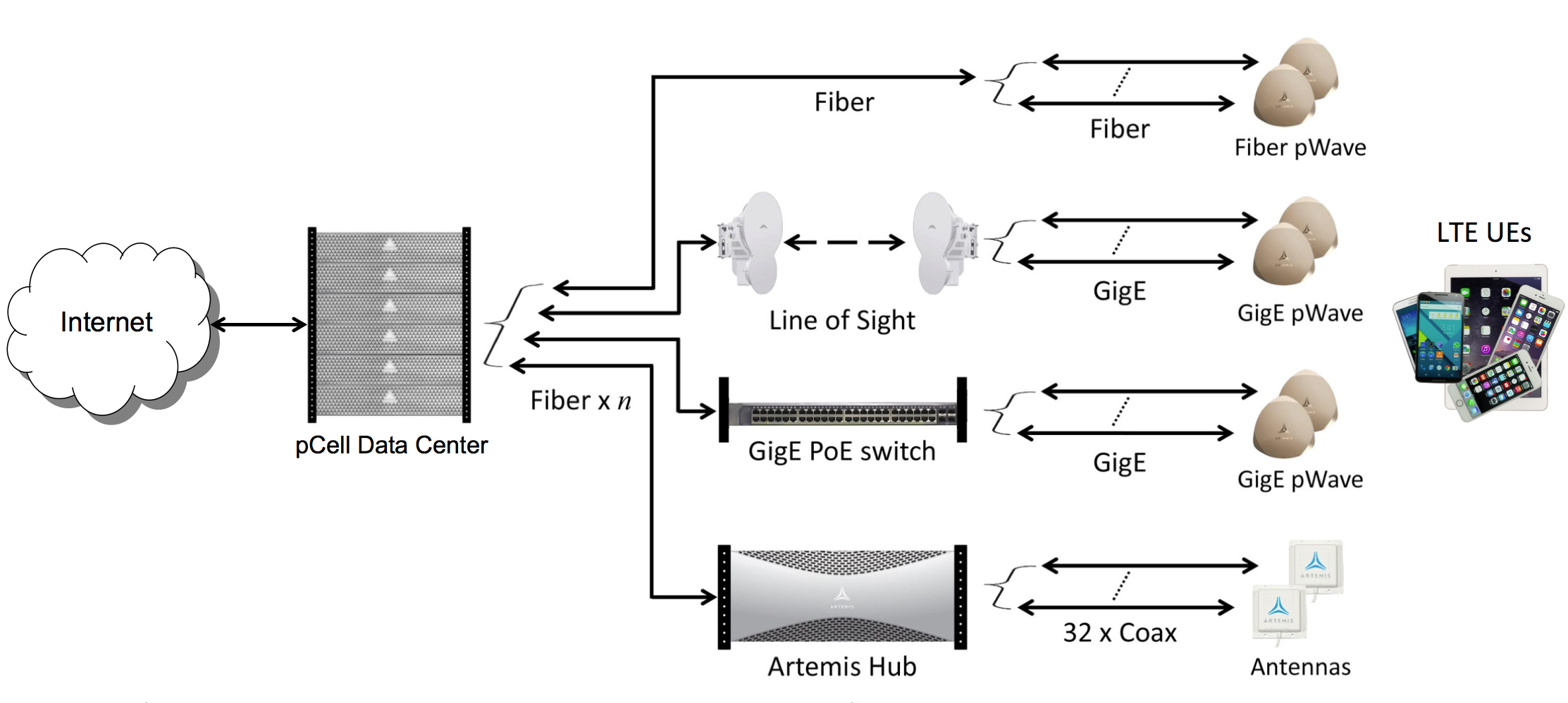}}}
	\caption{pCell hardware architecture.} 
	\label{fig:HW_arch}
\end{figure}

As illustrated in \fig{fig:HW_arch} the pCell system is composed of two parts: i) a GPP-based data center that implements the LTE protocol and pCell processing; ii)  a radio access network (RAN) including data switches and radio transceivers.
While the software running in the data center remains unchanged, the pCell RAN comes in different flavors so as to accommodate different deployment scenarios. A fiber fronthaul, which transmits a duplex of I/Q digital sample streams, is routed from the servers in the pCell data center and can be connected to:
\begin{enumerate}
    \item pWaves configured with a fiber interface;
    \item LOS radios that connect to pWaves configured for {1000BASE-T} (i.e., copper Gigabit Ethernet);
    \item the uplink port of a 1000BASE-T switch that connects to pWaves configured for 1000BASE-T;
    \item \label{ArtemisHub}an Artemis Hub composed of 32 pWave radios that connect to up to 32 antennas through coaxial cables.
\end{enumerate}

In this paper we present experimental results obtained with the RAN hardware configuration \#\ref{ArtemisHub}. The pWave radios can operate at any carrier frequency from 400 MHz to 4.4 GHz and are synchronized using 10MHz/PPS signals that can be retrieved from a GPS reference, or from an in-band timing signal itself slaved to a GPS reference. For the measurement results presented in this paper the GPP-based data center utilizes three off-the-shelf dual-processor Intel motherboards, two of them equipped with Intel Xeon E5-2687W v2 (8 cores @3.40GHz) and the third with Intel Xeon E5-2697 v3 (14 cores @2.60GHz).

\subsection{Software Architecture}

The pCell system is an SDR platform where all the functional blocks of the LTE protocol stack, from the gateway down to the physical layer, and all pCell processing have been implemented from scratch in C++ modules running in real-time on the GPP platform, without the need for specialized hardware such as DSPs, co-processors or FPGAs. This SDR implementation provides maximum flexibility, interoperability and portability. The software architecture is designed to minimize computational overhead, optimize throughput and provide stable and deterministic computational load behavior. The proprietary development framework provides a module-oriented environment with multi-threading/multi-core programming capability and support for real-time over-the-network operation.

For achieving the level of efficiency required to meet the real-time constraints of the LTE protocol and the pCell processing, the software modules are categorized in two classes: \textit{hard} real-time and \textit{soft} real-time. The \textit{hard} real-time modules implement tasks that must be completed within a fraction of the 1 ms LTE subframe (SF) duration. Completing these tasks on time is critical for maintaining the integrity of the pCell-synchronized LTE waveform and the stability of the data connection with the user equipment (UE). These operations include all of the pCell processing and most of the LTE physical layer functions (e.g., turbo coding/decoding, FFT, channel estimation/equalization) as well as some of the MAC layer features such as the PRACH procedure.
The \textit{soft} real-time modules implement functional blocks from the higher layers of the LTE protocol stack (e.g, RLC, PDCP, RRC, NAS, Gateways)  that are subject to time constraints in the order of multiples of an SF interval. The software architecture is built around this classification and, by using tools provided by the development framework, the system can balance the computational load and guarantee on-time execution of the critical \textit{hard} real-time tasks without missing the deadlines of the \textit{soft} real-time ones.

For each UE a set of computing resources and data structures, called a virtual radio instance (VRI), is allocated to instantiate an entire LTE protocol stack, thus forming the functional equivalent of a dedicated LTE eNodeB per UE.
A VRI is spawned as soon as a user begins the attach procedure and remains operational throughout the duration of the user's connection, maintaining its active state.
When a user detaches from the network, the VRI manager saves any relevant state and the VRI instance is released.
Through physical layer signal processing each VRI is associated with a volume of coherent signal (which was duly defined in Section \ref{subsec_Sinr}) around the antenna of the corresponding user.

As shown in Section \ref{subsec_Volumes}, such a volume of coherent signal is confined in spatial dimensions on the order of or smaller than the wavelength, resulting in a concurrent spatial multiplexing unit for each UE. Thus, every VRI has a concurrent full-bandwidth, independent physical-layer link with its associated UE, providing each UE with the experience of an unshared eNodeB delivering the full LTE channel bandwidth, regardless of number of VRI/UE links concurrently in the same spectrum.

\begin{figure} [!h]  
	\centerline{\resizebox{1.03\columnwidth}{!}{\includegraphics{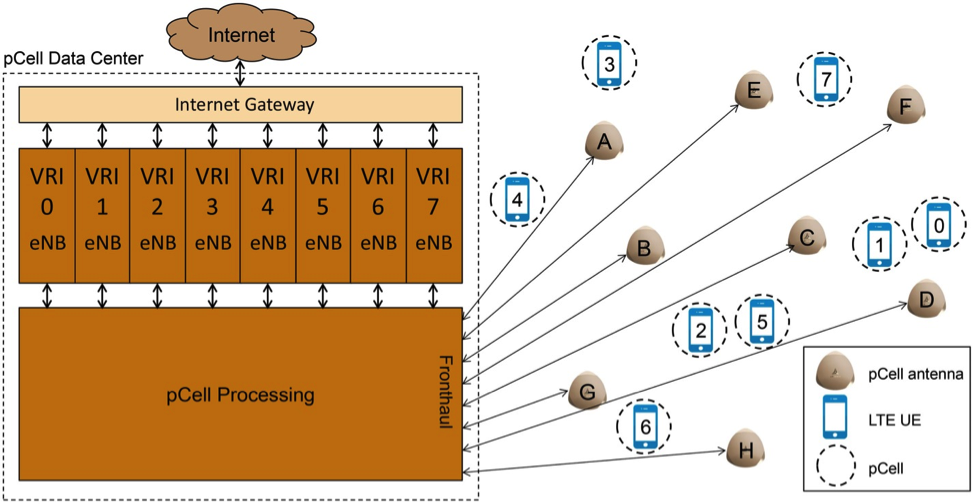}}}
	\caption{pCell software architecture.} 
	\label{fig:SW_arch}
\end{figure}

In \fig{fig:SW_arch} the spatial processing functions are denoted pCell processing and the volumes of coherent signals are conceptually depicted as dashed circles with numeric labels matching the labels of their associated VRIs.
Consequently, different VRIs can implement different protocols concurrently in the same spectrum. Some VRIs can implement LTE eNodeB while other VRIs can implement proprietary or standard protocols, for example, ones better suited for low-power ``Internet of Things'' devices. Both LTE and non-LTE devices will concurrently and independently operate in the same spectrum, with each device experiencing only the protocol from its own VRI.

\subsection{LTE Protocol and Frame Structure}
\label{sec_LTEProtocol}
\begin{figure} [t]  
	\centerline{\resizebox{1.04\columnwidth}{!}{\includegraphics{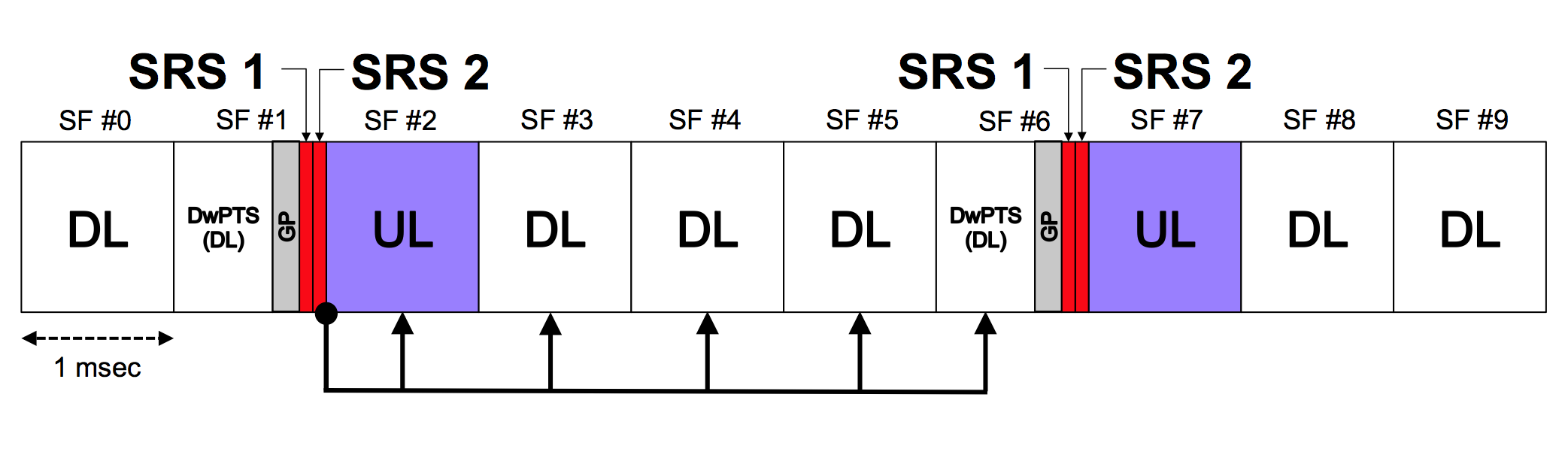}}}
	\caption{LTE TDD frame structure (TDD config. \#2 and Spc config. \#7)}
	\label{fig:lteTDDframe}
\end{figure}

The pCell system is compatible with off-the-shelf LTE devices. In LTE systems, the downlink (DL) uses orthogonal frequency division multiple access (OFDMA) and the uplink (UL) uses single-carrier orthogonal frequency division multiple access (SC-FDMA), with quadrature amplitude modulation (QAM) of different orders. The system bandwidth can range from 1.4 to 40 MHz. The frame structure changes depending on whether the frequency-division duplex (FDD) or time-division duplex (TDD) mode is used. The LTE frame is formed of 10 subframes (SF) of duration 1 ms each and  sub-divided into 14 OFDM symbols.
In TDD mode, the SF is either of type DL, UL or special (Spc). The Spc SF is placed between every DL and UL SF as in \fig{fig:lteTDDframe}, and consists of a downlink pilot time slot (DwPTS), a guard period (GP) and an uplink pilot time slot (UpPTS).
The GP allows for RF switching as well as timing advance to compensate for round-trip time of flight.
These parameters are broadcasted by the eNodeB for the UEs to set up compatible operation prior to attaching to it.

Within the LTE protocol, the DwPTS is used to transmit DL data and the UpPTS to send UL sounding reference signal (SRS) from the UEs, consisting of Zadoff-Chu sequences multiplexed over the entire system bandwidth. The SRS is used in current LTE networks to perform channel condition measurements for scheduling purposes. pCell uses the SRS to derive precise UL CSI and Doppler information from all UEs, while avoiding additional overhead compared to existing LTE networks. In 5 MHz bandwidth and with the TDD frame in \fig{fig:lteTDDframe}, every 5 ms up to 32 concurrent UEs transmit the SRS to the pWaves. In practical deployments for densely populated areas (e.g., stadium) users are typically stationary or nomadic, and different periodicities of SRS transmissions can be set depending on their speed as allowed by the LTE standard. For example, if the periodicity is set to 20 ms for all UEs, up to 128 concurrent users can be supported in the same bandwidth. Further, typical LTE deployments allocate 20 MHz blocks of spectrum which can be subdivided in four channels of 5 MHz each, thereby allowing up to 512 concurrent users with orthogonal SRS in the same coverage areas with 20 ms periodicity. In scenarios requiring even larger numbers of concurrent users, pCell employs spatial reuse schemes to increase further the number of orthogonal SRS transmissions.

The pWave radios synchronously receive the SRS signals, convert from RF domain to I/Q samples, which are sent to the pCell data center.  The samples are processed to produce accurate UL CSI estimates, which are used to derive accurate DL CSI by exploiting TDD reciprocity through proprietary signal processing algorithms. 
The UL/DL CSI is further processed to derive parameters for UL/DL spatial processing. Following the pCell processing operations on UL I/Q samples, conventional LTE physical (PHY) layer processing is applied to each user UL stream including equalization and turbo decoding to form a protocol data unit (PDU) passed to the MAC layer.
The UL data then proceeds up the LTE protocol stack within each VRI associated with a particular UE to eventually be sent to the Internet. The VRI also provides DL data through the protocol stack in the form of MAC PDUs scheduled for transmission in the DL SFs and DwPTS, processed according to conventional LTE PHY operations. The pCell processing then converts the $U$ streams of user DL samples into $N$ streams of pWave I/Q samples, which are finally transported to the pWave radios. The pWaves convert the I/Q samples to the RF domain and synchronously transmit the waveforms.

\section{Experimental Results}
\label{expResults}

\begin{figure}[t]
	\centerline{\resizebox{0.88\columnwidth}{!}{\includegraphics{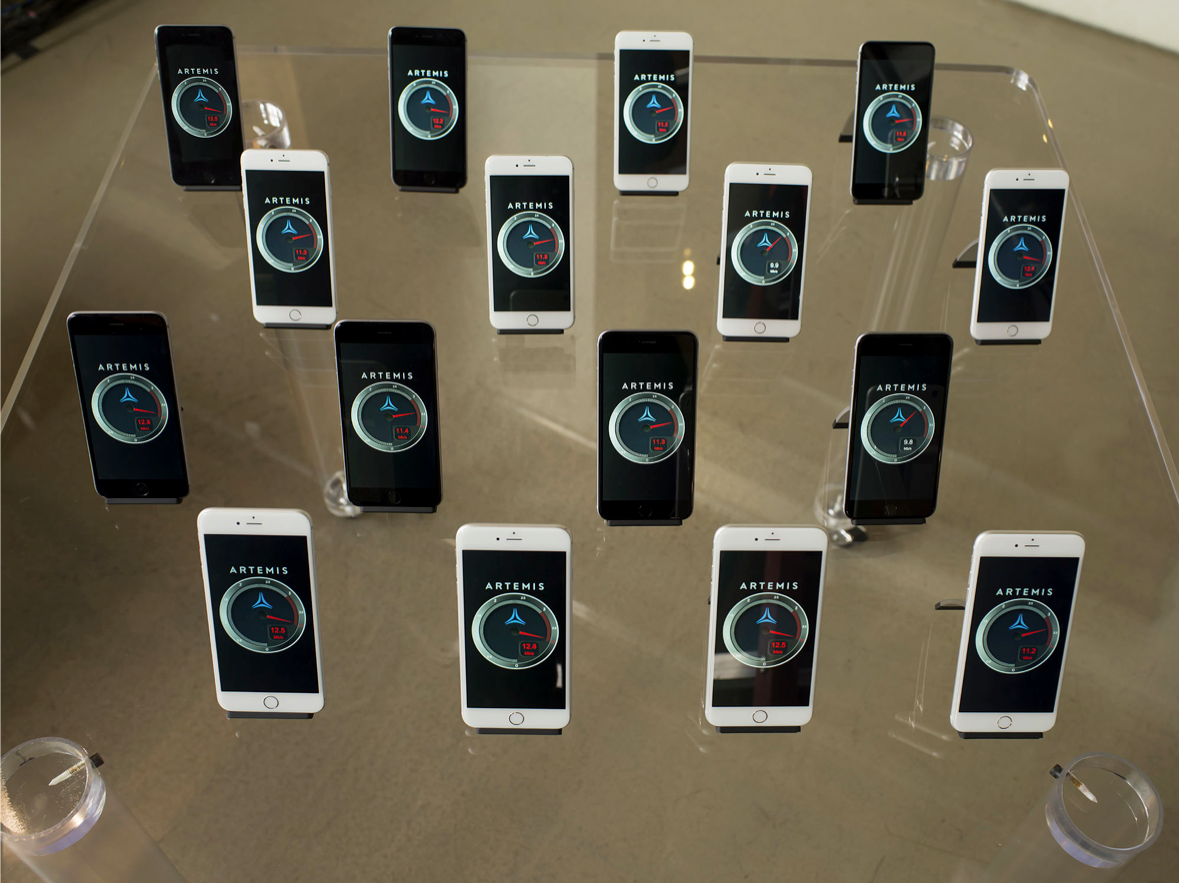}}}
	\caption{4G LTE devices densely packed in $1\,\textrm{m}^2$.}
	\label{fig:iPhones}
\end{figure}
\begin{figure} [b]
	\centerline{\resizebox{1.0\columnwidth}{!}{\includegraphics{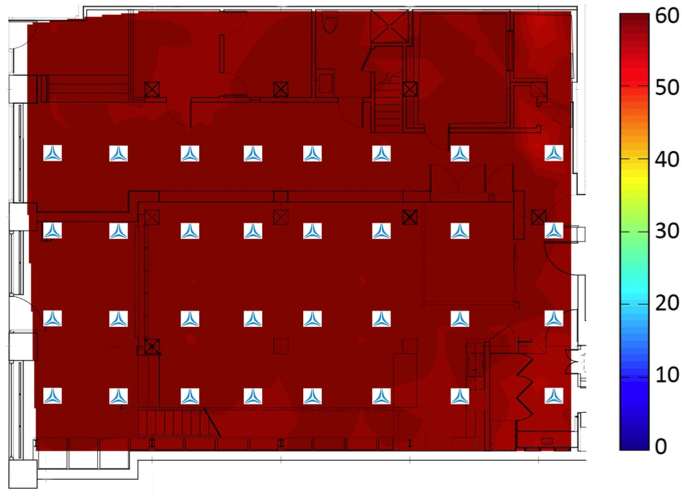}}}
	\caption{Downlink pCell SE (average = 59.3 bps/Hz, peak = 59.8 bps/Hz).} 
	\label{fig:DL_SE}
\end{figure}

pCell has been tested both indoor and outdoor, with detailed indoor testing completed thus far. The results presented in this paper have been obtained using the RAN configuration \#\ref{ArtemisHub} described in Section \ref{subsec_hardwareArchitecture}.
Although pCell supports arbitrary antenna placement, for the purpose of the SE testing, 32 antennas are placed in a regular grid with roughly 2.5 meter spacing, with aligned polarization, at a uniform height (except in low-ceiling corridors) and pointing downward, resulting in a mix of LOS and NLOS paths, some through walls and some through free space. 
The antennas are 2"x2" patch antennas with 8 dBi and HPBW = 75$^{\circ}$. Every antenna transmits a waveform with LTE-compliant spectral envelope generated in the data center with the time-domain frame structure in \fig{fig:lteTDDframe} and the following parameters:
\begin{itemize}
	\item Carrier frequency: 1917.5 MHz;
	\item System bandwidth: 5 MHz (with 300 OFDM subcarriers);
	\item Average transmit power per antenna: 1 mW;
	\item TDD configuration \#2 for a DL to UL ratio of 3:1;
	\item Spc SF configuration \#7. 
\end{itemize}

The pCell spectral efficiency (SE) is determined by measuring the aggregate SE of a group of users (unmodified iPhone 6 Plus devices are used for these measurements) all concurrently transmitting and receiving data from the pCell antennas in the coverage area. The iPhone 6 Plus devices are placed in a uniform pattern on a $1\,\textrm{m}^2$ plexiglass table as in \fig{fig:iPhones} moved throughout the coverage area in 75 cm increments.
The LTE MAC layer PDU throughputs are used to calculate the aggregate DL and UL SE at every location. The calculation of the LTE SE numbers presented in this paper are calibrated against the SE tables reported in \cite{3GPP-FS-2014}.

\begin{figure}[b]
	\centerline{\resizebox{1.0\columnwidth}{!}{\includegraphics{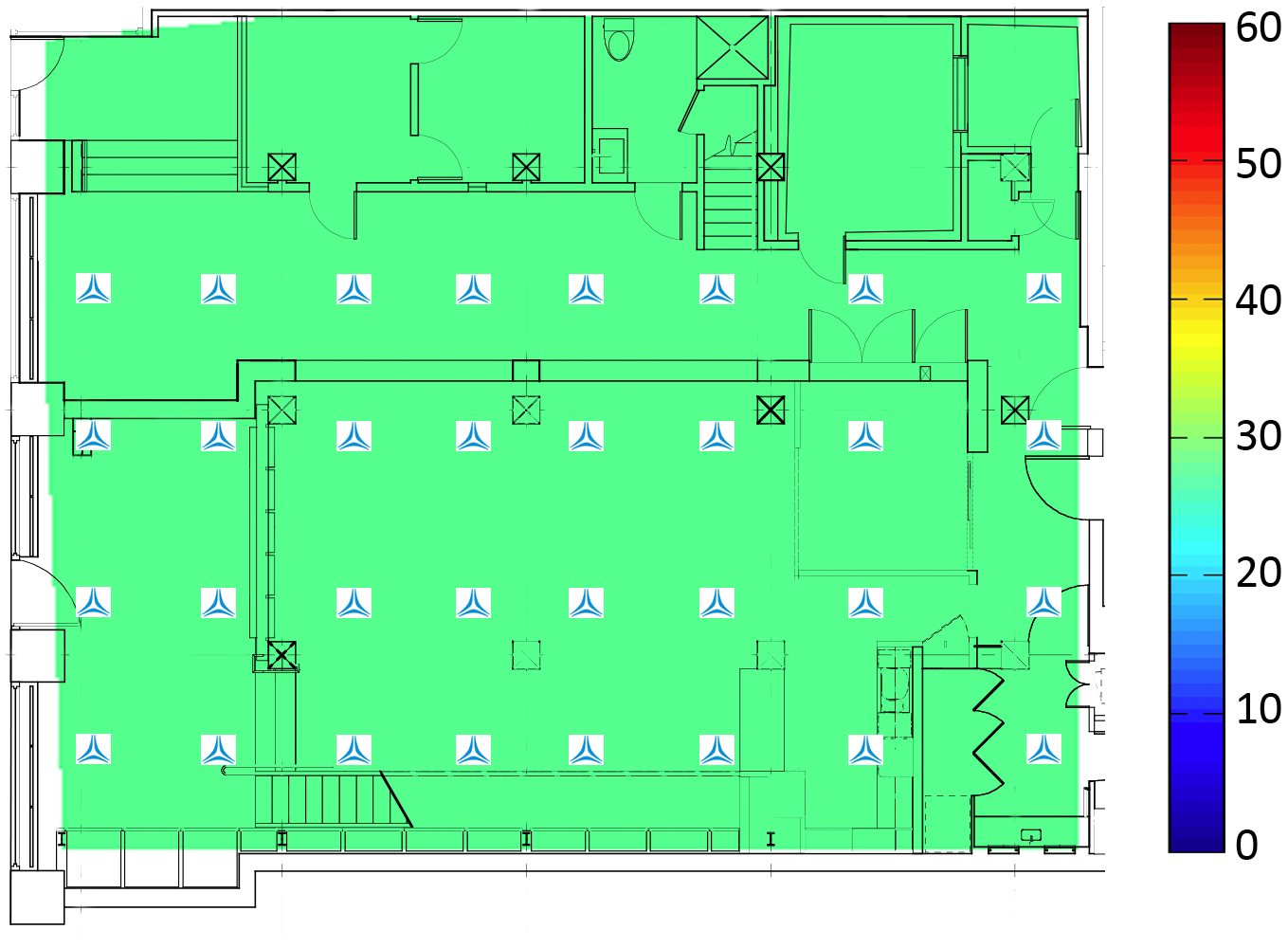}}}
	\caption{Uplink pCell SE (average = 27.5 bps/Hz, peak = 27.5 bps/Hz).}
	\label{fig:UL_SE}
\end{figure}

Heat maps of the aggregate DL and UL SE for the 16 iPhone 6 Plus devices throughout the coverage area are shown in \fig{fig:DL_SE} and \fig{fig:UL_SE}, respectively, along with the layout of 32 antennas (white squares with the blue Artemis logo).
The average aggregate DL SE across all locations is 59.3 bps/Hz, with peak of 59.8 bps/Hz (corresponding to an aggregate DL throughput of approximately 200 Mbps) and 5\% outage \cite{ITU-2008} of 58.1 bps/Hz mostly due to locations in the upper right corner (obstructed by several walls). At its DL SE peak, every UE receives data using LTE modulation and coding scheme (MCS) 28 which corresponds to a 64-QAM OFDM modulation with FEC coding rate of 0.9.  
\fig{fig:UL_SE} shows the aggregate UL SE is consistently at peak of 27.5 bps/Hz (corresponding to an aggregate UL throughput of 25 Mbps) throughout the entire coverage area. Every UE transmits data using MCS 20 (i.e., 16-QAM SC-FDMA modulation with FEC coding rate of 0.7), which is the maximum UL MCS supported by LTE UE category 4 chipsets \cite{3GPP-UECAT-2015} used by the iPhone 6 Plus.

Because pCell processing results in high SINR throughout the coverage area, in almost all locations the SE is limited by the maximum MCS supported by the iPhone 6 Plus (MCS 28 for DL and MCS 20 for UL). Future LTE devices are expected to enable higher order modulation schemes (e.g. 256-QAM \cite{3GPP-CQI-2014}), which pCell will be supporting given the large SINR margin it provides.

\section{Conclusion}
\label{sectConclusion}

We presented pCell, a multiuser wireless system that enables cooperation between distributed transceivers to achieve large multiplexing gain while foregoing cellularization. By forming individual volumes of coherent signal with high SINR around every user's antenna, pCell creates independent spatial links to multiple users, thereby providing over an order of magnitude increase in spectral efficiency compared to existing technologies. We introduced an analytical framework based on a geometrical channel model to describe the volumes of coherent signal. Then we reviewed hardware and software architecture of pCell as implemented within the LTE protocol. Finally we demonstrated high downlink and uplink spectral efficiency achieved with pCell in practical propagation conditions.


\bibliographystyle{IEEEtran}
\bibliography{references,giuseppe-refs}

\end{document}